# Re-designing knowledge management systems :
## *Towards user-centred design methods integrating information architecture*

Carine Edith TOURE[1,2], Christine MICHEL[1,2] and Jean-Charles MARTY[3]

[1]*Université de Lyon, CNRS*
[2]*INSA-Lyon, LIRIS, UMR5205, F-69621, France*
[3]*Université de Savoie, LIRIS, UMR5205, France*
*{carine-edith.toure, christine.michel, jean-charles.marty}@liris.cnrs.fr*

Abstract: The work presented in this paper focuses on the improvement of corporate knowledge management systems. For the implementation of such systems, companies deploy can important means for small gains. Indeed, management services often notice very limited use compared to what they actually expect. We present a five-step re-designing approach which takes into account different factors to increase the use of these systems. We use as an example the knowledge sharing platform implemented for the employees of Société du Canal de Provence (SCP). This system was taken into production but very occasionally used. We describe the reasons for this limited use and we propose a design methodology adapted to the context. Promoting the effective use of the system, our approach has been experimented and evaluated with a panel of users working at SCP.

## 1 INTRODUCTION

Knowledge books refers to a class of knowledge management systems (KMS) that are built according to the MASK method (Aries, 2014). They use models of the activity field and are widely used within companies because they promote easy search, understanding and use of information. They also help structuring information and documents with hypertext links. Nevertheless, these KMS are efficient only under some terms like effective reading and updating of documents. (Prax, 2003) introduces in his book several reasons of knowledge management (KM) process failures in companies. We can find functional problems (e.g. the system is too complex, the system does not address users' needs, there is poor organization of information), management problems (e.g. there are inadequate or non-existent change management strategies), and even human problems (e.g. the KMS type is inadequate to the cultural context of the company). Some other issues are directly related to methods used to design KMS. Indeed, KM designers first strive to properly structure and define core concepts and contents in order to describe them. It is only in a second step that they focus and with less expertise, on interfaces design and terms of interactions. Thus, aspects like ergonomics, content organization and indexing, KMS integration with other information systems are much less considered. This work proposes a design methodology to develop KMS that meet more effectively the needs of demanding users in businesses. We take into account the fact that the corporate knowledge is already formalized in models to produce knowledge books, our proposal thus focuses on re-designing an existing KMS. Our method integrates, in particular, user-centred design processes and information architecture. We tested our methodology in a real context of use so we can fetch elements of evaluation and feedbacks.

*Société du Canal de Provence* (SCP) is a hydraulics services company located in the Provence Alpes Côte d'Azur French region. This company agreed to participate to our study; since 1996, SCP has massively invested in a knowledge book named ALEX (*Aide à l'EXploitation*). For more than ten years of production, the management services of the company observed a weak use of ALEX, despite the fact that the collaborators are really aware of the usefulness of such a system. A preliminary study was carried out to identify ergonomics-related issues, difficulties in updating information and access processes. This context was particularly representative of design problems that one can encounter when designing KMS. The general issue we are trying to solve in this paper is how to re-design KMS to promote their use. We propose studying how to combine knowledge engineering design methods, human-machine interaction design methods and architecture of information. This article is organized as follows: in the next section, we introduce a number of methods that can be used for designing; then we propose our methodology. In section four, we describe the implementation and how ALEX has evolved in SCP. We measured the effectiveness of the method by conducting a user survey. Section five draws a conclusion and identifies areas for future research.

## 2 REVIEW OF THE LITTERATURE

Research work in knowledge engineering has enabled the development of several methods to enhance and capitalize workers' expertise in the company. A more suited way to implement these systems in an industrial context would be to consider methodologies and techniques taking into account the needs and expectations of employees in terms of collaboration and communication. The fields of Human-Computer Interaction (HCI) and Information Architecture (IA) document the literature on these methods.

### 2.1 KMS design using pioneering methods

Design methods in knowledge management aim to formalize a set of procedures, skills, life skills, etc.., which structure the activity in a way to make them re-usable. The KOD method proposed by Vogel in 1988 (Ermine, 2008), focuses on modelling expert knowledge from his free speech on his activity. It consists of at least three steps: position of the subject, interview and analysis. The MEREX method (Prax, 2003) developed at Renault offers to produce experience feedback in the form of sheets. This sheet database is the KMS. An animation procedure including three actors is then implemented to keep the records up to date. The editor, not necessarily an expert, produces sheets, while the validator that checks the correctness of the contents must be a domain expert; the manager supervises the system. Methods of knowledge engineering provide tools for the formalization and capitalization of knowledge. They mainly involve professionals in the field, essentially focused on the formalization of knowledge and do not consider the user who is supposed to interact with the system. This positioning may hinder the effective use of knowledge, as illustrated in the introductory section.

### 2.2 The active participation of users in the design process

The HCI field aims to study, plan and design the modalities of interaction between the user and the computer. The challenge is not only to produce useful, usable and acceptable interactions but also to improve the user experience according to its value system, its business context and objectives. In fact, the user experience is defined as the set of a user's perceptions in a situation of interaction with a product (Garrett, 2011). These perceptions determine the success or failure of a product. In (Nielsen, 1993), the author defines the User-Centred Design (UCD) as a philosophy and a design approach where the needs, expectations and characteristics of end users are taken into account at every development stage of the product process. This task can be complex since users do not generally know what they want. (Mathis, 2011) presents some approaches to better understand the expectations of users and implement them in a more effective way. The methods of Twinning (also called Job Shadowing) and Contextual interviews are examples of these approaches. The UCD refines the understanding of the needs, possibilities and limitations of the user in his/her activity and regarding a technological proposal. It allows designing more responsive systems for human and organizational contexts where they are used.

### 2.3 Information architecture methods

As KMS can be considered as a class of information systems, we sought what methods could be used to define the way people access to information in such systems. IA is a discipline that seeks to define how to present information in the most appropriate manner, depending on the future user and on the context of use. The goal of IA is to define the form of presentation that will make the most usable information in terms of understandable and usable. Thus, a good information architecture must be searchable (the user is not confused), coherent (semiotics adapted to the context of use), adaptable, simple (just enough information presented), able to make information recommendations (Resmini & Rosati, 2011). Different methods are proposed to achieve these objectives. (Resmini & Rosati, 2011) indicate taxonomic methods (identification of hierarchical relationships and semantic similarity between concepts), methods of Sorting Cards that are used to identify and organize super categories and intermediate categories, representing the major classes of users' needs. (Garrett, 2011) proposes a more comprehensive conceptual framework for structuring the design, especially suited for Web applications. Its main feature is to separate and coordinate the design features of the product and the information it must take into account. This method is based on five steps: strategy, scope, structure, skeleton and surface. These steps describe how to shift from abstract elements of the product design to more concrete elements. In order to structure information, Garrett proposes to use either a top-down approach (based on the needs and objectives of the system for organizing information) or a bottom-up approach (use categories and subcategories to organize the information).

## 2.4 Summary

Conventional knowledge management methods prove less effectiveness because they often lead to KMS that do not fit user's expectations. We observe companies adopting new Web 2.0 technologies such as social networks. This is explained by the fact that these new modes of exchange and communication are becoming more and more casual in users lives and they provide socialization platforms to support business initiatives for knowledge seeking and sharing. Moreover, the techniques of participatory design and structuring of information provide the means for designers to stay close to target users. We propose in the next section, a methodology to re-design a classical KMS, from a knowledge book to an enterprise social network.

## 3 A FIVE STEP APPROACH

Companies are already aware of the importance of formalizing the corporate knowledge and building KMS. Most of them have already invested in these strategies but could get neither a really usable system nor a system accepted by the collaborators. This is why we propose a re-designing approach based on an existing knowledge capitalization. Our approach is inspired by the conceptual framework of (Garrett, 2011) and consists of five phases operating tools and methods from the KM, the user-centred design and information architecture. The steps are cyclic and may overlap if needed. At the end of a cycle, an evaluation of the resulting prototype is carried out and can lead to a redefinition of the previous choices.

## 3.1 Needs analysis

In this step, we propose to analyse the initial situation of the KMS, the users' needs and the system objectives. This information gathering allows us to work on capitalized knowledge, the business context and the expectations of the users and of the company. The objective is to observe difficulties of the users in front of the existing KMS and find appropriate solutions. For a knowledge book, it is interesting to observe users' reaction in front models of organization, media or forms of navigation. Concerning the other aspects, an immersion in the industrial environment is required; matching methods and contextual interviews are appropriate means to understand the environment of use and the needs (Mathis, 2011). In addition, the method of mind-mapping (Prax, 2003) can help to better understand the business vocabulary through games characterization. These studies are useful to classify the important business concepts into categories and super categories that will help us to structure the information in step 3.

## 3.2 Definition of the new KMS

This step is used to define the features for managing knowledge (information sharing, information seeking, collaboration, learning) which meet the needs. The case of SCP led us to consider the target KMS as an enterprise social network (ESN). Companies are increasingly interested in easy sharing and exchanging of information via ESN (Stocker & Müller, 2013) and look for ways to exploit them. (Zammit & Woodman, 2013) justify the effectiveness of ESN in knowledge management by carrying out an analysis based on the SECI model of Nonaka and Takeuchi. However, there is no heuristic describing the best way to operate an ESN as a KMS. In this stage of our method and to refine the design choices, we recommend working in parallel on the definition of roles and features because these elements are essential in ESN. The definition of roles can be done using the MEREX method (Prax, 2003). Then one must specify, by methods such as focus groups, the usefulness and the format of features like members directory, news, statistics, newsletter, photo album, dialogue groups, events, polls, discussion forums.

### 3.3 Design of the information structure

This step addresses the specification of interaction formats adapted to the features and information architecture of the system. This phase is closely related to the results of the analysis in Phase 1. It consists in designing patterns of interaction and models to structure information that are familiar to users. As the forms of interaction of the target CMS are already pervasive in users' habits, we recommend keeping them. Structuring the content on the other hand, should be investigated. Among the two approaches proposed by Garett and presented above, we prefer the bottom-up approach. Indeed, a method of sorting card identifies the key concepts as well as their different facets but also different forms of information to be exploited.

### 3.4 Design of the skeleton

This step is used to design the main functional areas and how they are interconnected. The user-centred method that we recommend in this step is the definition of personas (Boucher, 2007). Personas are virtual characters that correspond to the end users, they allow designers to really get into the skin of end users and make proposals that best fit them. They are important to specify the navigation schema connecting the different functionalities.

### 3.5 Visual design

The general graphical appearance and textual fonts will be determined in this phase. We recommend setting the visual design in accordance with the Charter of communication and graphical design of the company to maintain familiarities with interfaces. Moreover, improvements in usability can be made if necessary. One possibility is to stage the main use cases, and discuss with the users on proposed models. We therefore recommend a prototyping approach which allows the improvement of the visual design.

The last two steps of our methodology, design of the skeleton of the platform and visual design, are highly sensory and will impact the user experience. The designer will have to be particularly careful when developing the platform in terms of ergonomics; He will also be able to develop motivators, all in accordance with user's expectations. This may require recycling regularly through these steps.

## 4 IMPLEMENTATION

SCP offers an application context for the implementation and evaluation of our methodology. It is specialized in services related to the treatment and distribution of water for companies, farmers and communities. It employs a significant number of persons, operators responsible for the maintenance of hydraulic structures also called infrastructures (e.g. canals, pumping stations, water purification stations). Operators work in a territory which is divided into ten geographical areas called operation centres. SCP has developed a KMS, ALEX which has the same features as a knowledge book. It contains information on the activities of agents of the SCP and makes this information accessible through experience sheets in html format. Faced with the problem of limited use of ALEX by the operators, we proposed to SCP to redesign their system using our methodology. The re-design was done with a working group of twelve people that are representative in terms of function and competence of all future users of the system. It lasted five months and six meetings allowed to monitor the project.

### 4.1 Deployment of the methodology

#### 4.1.1 Needs analysis

The study of the initial system showed that the information filled in on business procedures was generally of good quality, which is fine since we haven't had to rework on the formalization of the knowledge itself. By cons, this phase allowed us to detect that the original system did not meet the basic employees' needs, what explained the limited use of the system. Interactions with the panel during this phase allowed the users to express their wish to have an accessible platform both in the office and outside, that is easy to learn, but effective, which reduces the time for entering information, which facilitates data search and allows exchanges between employees. In addition, our intervention focused on the type of the KMS, we went about designing a website for ESN.

#### 4.1.2 Specification of the ESN

During focus groups, the following roles have been defined: the administrator, the validator, the contributor and the commentator. Employees, depending on the role they have in the system, are more or less involved in the animation of the platform and content validation. These roles helped develop access features for system security levels and promote empowerment of actors. We proposed to collaborators to integrate in the ESN, features such as the submission and publication of pre-formatted forms, submitting comments on the experience sheets, news from the operation centres, photo albums, discussion forums, features to moderate submissions based on roles. After discussion, it was decided that the submission is made by the contributor under moderation of the validator or the administrator. It is done using forms with large areas of open writing because the activity is too complex to be defined by a structure; employees without any role distinction communicate by commentaries left on the sheets. These comments serve to convey the appreciation of the reader regarding the record, as its content is a good idea to generalize or if there is a need for additional information; research content is natural language or keywords; photos of the album will be indexed according to the operating structures and equipment in order to facilitate research centres.

### 4.1.3 Design of the information structure

In this stage, we identified informational patterns related to business concepts: operations done by the collaborators on infrastructures are described in experience sheets, descriptions of equipment, operating instructions, interventions of process type, alarms, etc. This has resulted in the definition of eleven types of sheets: presentation, equipment, set operating instructions, hydraulic diagram, spreadsheet, process, operation, alarm, contract / contact and detail. Each item is described by a presentation sheet that can be associated with one or more sheets of the other types. Each sheet is presented as a form formatted according to its type-specific metadata (type of equipment for equipment sheets or instruction name for operating instruction sheets), by cons, the content of fields is open written.

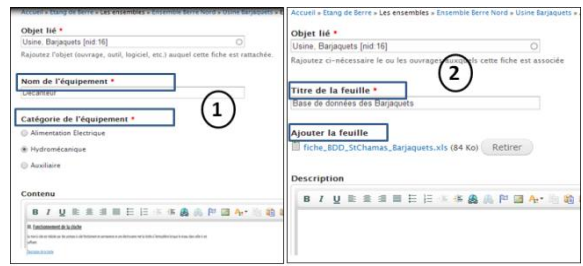

Figure 1: View of two types of sheets. Form 1 is an equipment sheet and the 2$^{nd}$, a worksheet; all the fields differ depending on the type.

### 4.1.4 Skeleton design

This phase was animated by discussions on the basis of a proposal of a skeleton made with a content management system (CMS) named Drupal. The use of a CMS allowed us to accelerate the development and modifications of the prototype according to the users' feedback. In the light of the opinions that have been collected, we made the proposal to reorganize the new system in different functional areas gathering the main features of ALEX that are information, submission, navigation and communication. The content is classified in accordance with the operational centres of the users and the types of sheets.

### 4.1.5 Visual design

On the visual aspects of the site, we started with a proposal of "theming" and, according to different users' feedback, we adapted the different choices. In this phase, we had to get into the skin of each user type and depending on the use case, to make choices that facilitate the use of the tool. As examples, we can cite the search area that has been enlarged and positioned prominently at the top right corner, the default font for the fill of sheets that has been standardized to Arial, size 12, or the area for data entry that has been enlarged.

### 4.1.6 ALEX+

This methodology has enabled the delivery of a new version of ALEX named ALEX +. Features have been proposed in order to improve the user experience in terms of complexity and duration, during the activity of corporate knowledge. An assessment has been made and is detailed in the next section.

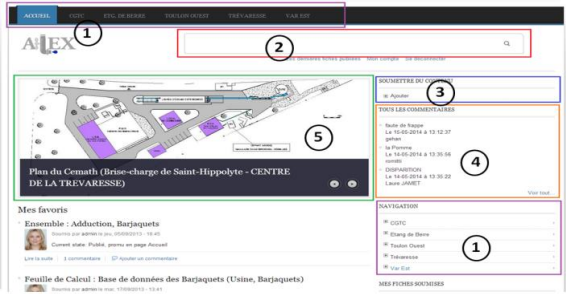

Figure 2: View of the ALEX+ front page. Zones 1 present the different types of navigation by tabs at the top or a menu block on the right. Zone 2 shows the search form and in zone 3 we positioned the block content submission. The regions 4 and 5 show a slideshow of the comments on sheets and the photos posted.

## 4.2 Evaluation

The ergonomic cognitive psychology assumes three dimensions for systems evaluation: systems that are usefulness (does the system meet the user's needs?), usability (how does he respond to the user's needs?) and acceptability (is it acceptable?). To validate our approach, we defined criteria to measure the level of utility, usability and acceptability of our prototype.

### 4.2.1 Evaluation criteria

We used the quality of the system and information, utility, usability and satisfaction. They were chosen because are they are emblematic of the success of the technology acceptance according to different models. TAM (Davis, 1993) and the UTAUT models (Venkatesh et al., 2003), consider the perceived usefulness and perceived ease of use as basics of attitudes and behaviours of users. The ISSM model (Delone, 2003) considers the quality of the functionalities, the information and services provided by the system as determinant for intention and effective use of the system. ISO 9241-11 model assumes that usability (effectiveness, efficiency and satisfaction) is a necessary condition for use (Février, 2011). According to these reference models, we constituted a summary model that describes how the process of acceptance should occur. The following figure presents this summary model with the evaluation criteria we used to form the questionnaire and evaluate our methodology.

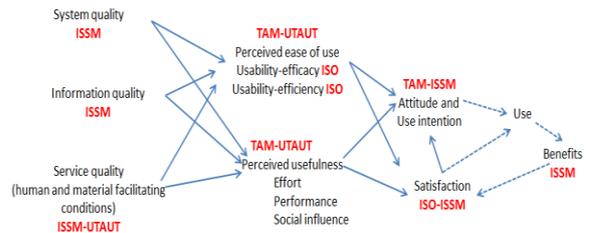

Figure 3: Summary model for acceptance of systems

### 4.2.2 Data gathering

The evaluation was conducted on the entire user panel, namely the 12 members of the working group. This was about completing a questionnaire organized into nine sections describing the nine factors listed in the previous model: System Quality (QS), Quality of information (QI), perceived Ease of use (FUP), Attitude toward the use of the application (ATT), Intention to use (IC), Service Quality (How is the system proposed by the SCP?), professional perceived Usefulness related to the use of the application (UPR), Personal perceived Usefulness related to the use of the application (UPE) and Satisfaction (SA). Each section contains a series of 4 to 6 questions (Touré, 2013).

### 4.2.3 Results

The results (cf. Table 1) show a majority of positive answers to the questions concerning the quality of the system and information, the perceived usefulness, acceptability and satisfaction. Users are generally satisfied and feel comfortable with the idea of using a social network as a platform for collaboration and sharing. This helps us assume that we will notice an increase of the system use when it will be put in production in the company. One can therefore note that a proportion of users do not perceive the professional and personal usefulness of the system. It can be explained by the fact that they are experts in their field and that the use of the tool does not necessarily provide them with an improvement of their skills. Users with this profile will tend to submit naturally but in long term they can find less or no motivation to use the system. Initial acceptance of KMS does not guarantee continuous and sustainable use (He & Wei, 2009). This issue will be part of the study in future works.

|        | QS and QI |      | UP   |      |      | ATT and IC |      |      |
|--------|-----------|------|------|------|------|------------|------|------|
|        | QS        | QI   | UPR  | UPE  | FUP  | ATT        | IC   | SA   |
| PA(%)  | 96,3      | 83,3 | 64,4 | 66,7 | 88,9 | 85,2       | 71,1 | 86,1 |
| NA(%)  | 0         | 2,8  | 26,7 | 22,2 | 0    | 0          | 17,8 | 0    |
| A(%)   | 3,7       | 13,9 | 8,9  | 11,1 | 11,1 | 14,8       | 11,1 | 13,9 |

Table 1: Results. PA: Positive answers; NA: Negative answers; A: Abstentions.

## 5 CONCLUSION

Conventional methods of KM are mainly based on the content and the formalization of knowledge. Considering these limitations, we proposed a user-centred approach to redesign KMS. Our methodology combines HCI methods and IA. Indeed, KM engineering methods are used to identify and formalize corporate knowledge. They are limited because they generally pay less attention to users' point of view regarding the exploitation of the knowledge produced. HMI methods are mainly used to ensure that systems are useful, usable and acceptable. The information architecture is used to ensure that knowledge presented in KMS follows structures that make the most sense for users and their organizational context.

We implemented our approach in SCP and as a result we obtained a prototype, ALEX +. ALEX + was evaluated by a panel of users. It shows that collaborators are generally satisfied with the proposals that were made in the final system and will tend more to use it. We can however identify a couple of limitations in our approach. Firstly, the limited number of participants in the workgroup allows us to only have the viewpoints of a small part of the actual user population; an assessment of a larger amount of people in SCP and also in other companies would help us have a better insight of the impact of our methodology on the KMS use in the company. Secondly, an ideal experimental approach would be to do a comparative evaluation of our methodology with others proposed by literature in the domain of design of corporate KMS. These points are planned for future work.

More generally, with our approach, we can just have an overview of the users' intentions but not of the effective use. Our method is not robust enough to ensure effective use; it focuses on initial acceptance of the system but not on his continuous use. A KMS is really useful if users effectively consult or add new content, discuss or comment updates, which happens when they master the system. This form of capitalization, which we call sustainable, requires implementation of other features in the system. This stage corresponds to the sensory design stage which we did not particularly emphasize in our approach.

We believe that metacognitive assistance features like indicators of awareness may be useful (Marty & Carron, 2011). Indeed, by proposing activity indicators, we can promote a reflexive dynamic of learning by user self-regulation processes (George, Michel, & Ollagnier-Beldame, 2013). For example, users by visualizing the impact of their contribution on other actors in the company may be more motivated to use the system. Conversely, by identifying the comments that were made on experience sheets related to their professional field, they may become aware of new procedures or changes in business practices and thus increase the credit given to the developed tool. As such, comments could be seen as a recommendation to consult. We plan to implement these new features by analysing traces of activity (Karray, Chebel-Morello, & Zerhouni, 2014). These traces provide much more diagnostic of use by sector and functionality. Our future work will therefore seek to identify, still with an incremental approach, which indicators and interaction modalities may be most suitable. Phases 4 and 5 of our method are mainly concerned; the design that affects the sensory and user experiences.